# Test for penetration in Wi-Fi network: attacks on WPA2-PSK and WPA2-Enterprise

Tamara Radivilova
Dept. of infocommunication engineering
Kharkiv National University of Radio Electronics
Kharkiv, Ukraine
tamara.radivilova@gmail.com

Hassan Ali Hassan
Dept. of infocommunication engineering
Kharkiv National University of Radio Electronics
Kharkiv, Ukraine
tameer2009@gmail.com

*Abstract—* **In this work the wireless networks security algorithms were analyzed. The fundamentals of the WPA and WPA2 safety algorithms, their weaknesses and ways of attacking WPA and WPA2 Enterprise Wireless Networks are described. Successful attack on the WPA2-PSK and WPA2-Enterprise was carried out during the performance of work. The progress of this attack and its results were described.**

## I. INTRODUCTION

The problem of protecting corporate data every year is more relevant. More and more critical data is transmitted over wireless networks, and information security (IS) increasingly depends on the skills of IT professionals. Many organizations and individuals use wireless local area networks (WLANs) as an irreplaceable addition to traditional wired LANs. WLANs are necessary for mobility, special networks and for access to hard-to-reach places. Many modern devices that we use (smartphone, tablet, laptop, router, TV), can work with wireless networks Wi-Fi. The most common at the moment is the IEEE 802.11i standard.

Any interaction between an access point (network), and a wireless client, is built on: authentication - both the client and the access point are presented to each other and confirm that they have the right to communicate among themselves; encryption - which algorithm of scrambling transmitted data is used, how the encryption key is generated, and when it is changed [1,2].

A lot of attention is given to Wi-fi network security. However, networks can be tested for security. The purpose of this work is to implement attacks on the Wi-fi network protected by the protocols WPA2-PSK and WPA2-Enterprise.

## II. SECURITY OF WI-FI NETWORK

The parameters of the wireless network, primarily its name (SSID), are regularly announced by the access point in the broadcast beacon packets. In addition to the expected security settings, QoS wishes, 802.11x parameters, supported speeds, information about other neighbors, etc. are transmitted. Authentication determines how the client is presented to the point. Possible options: open – so-called open network, in which all connected devices are authorized immediately; shared – authenticity of the connected device must be verified with a key/password; EAP – authenticity of the connected device must be verified by EAP with an external server [1,3,4].

The openness of the network does not mean that anyone can work with it with impunity. To transmit data in such a network, it is necessary to match the encryption algorithm used, and, accordingly, to correctly establish the encrypted connection. Encryption algorithms are as follows: none – no encryption, the data is transmitted in clear text; WEP – is a cipher based on the RC4 algorithm with different lengths of a static or dynamic key (64 or 128 bits); CKIP – proprietary replacement of WEP from Cisco, early version of TKIP; TKIP – improved WEP replacement with additional checks and protection; AES/CCMP – is the most advanced algorithm based on AES256 with additional checks and protection [3,4].

The combination of Open Authentication, No Encryption is widely used in guest access systems such as providing Internet in a cafe or a hotel. To connect, you only need to know the name of the wireless network. Often, this connection is combined with an additional check on Captive Portal by redirecting the user's HTTP request to an additional page where you can request confirmation (login-password, agreement with rules, etc.). WEP encryption is compromised, and it can not be used (even in the case of dynamic keys). Widely encountered terms WPA and WPA2 determine, in fact, the encryption algorithm (TKIP or AES) [2]. Due to the fact that client adapters support WPA2 (AES) for quite some time now, it makes no sense to apply encryption using the TKIP algorithm.

The difference between WPA2 Personal and WPA2 Enterprise is where the encryption keys used in the mechanics of the AES algorithm come from. At home and in small offices, PSK (Pre-Shared Key) is usually used - a password of 8 characters. This password is the same for all, and is often too simple, so it is vulnerable to selection or leaks (firing an employee, a missing laptop, an inadvertently glued sticker with a password, etc.) [5]. Even the latest encryption algorithms when using PSK do not guarantee reliable protection and therefore are not used in serious networks. Corporate solutions use a dynamic key for authentication, which changes each session for each user. The key can be updated periodically during a session using an authorization server - usually a RADIUS server.

Using WPA2 Enterprise requires a RADIUS server on your network. At the moment the most efficient are the following products:

Microsoft Network Policy Server (NPS), formerly IAS - configured via MMC, is free, but you need to buy a Windows;

Cisco Secure Access Control Server (ACS) 4.2, 5.3 - configurable through the web interface, feature-rich, allows for the creation of distributed and fault-tolerant systems, is expensive;

FreeRADIUS - free, configured with text configurations, in management and monitoring is not convenient.

The EAP protocol itself is container, that is, the actual authorization mechanism is given for the purchase of internal protocols. At the moment, any significant spread has been received by the following [2,4]:

- EAP-FAST (Flexible Authentication via Secure Tunneling) - developed by Cisco; Allows you to authorize by login-password transmitted inside the TLS tunnel between the supplicant and the RADIUS server

- EAP-TLS (Transport Layer Security). Uses a public key infrastructure (PKI) to authorize a client and server (a grantee and a RADIUS server) through certificates issued by a trusted certificate authority (CA). Requires issuing and installing client certificates for each wireless device, so it is only suitable for a managed corporate environment. The Windows certificate server has facilities that allow the client to generate a certificate on its own, if the client is a member of the domain. Blocking a client is easily done by revoking its certificate (or through accounts).

- EAP-TTLS (Tunneled Transport Layer Security) is similar to EAP-TLS, but the client certificate is not required when creating the tunnel. In such a tunnel, similar to the SSL connection of the browser, additional authorization is performed (by password or otherwise).

- PEAP-MSCHAPv2 (Protected EAP) - similar to EAP-TTLS in terms of the initial establishment of an encrypted TLS tunnel between the client and the server that requires a server certificate. In the future, such a tunnel is authorized by the known protocol MSCHAPv2

- PEAP-GTC (Generic Token Card) - similar to the previous one, but requires one-time password cards (and corresponding infrastructure)

### III. METHODS OF HACKING WI-FI NETWORK

For encryption based on WEP, only the time to browse IV is required, and one of the many freely available scan utilities. For encryption based on TKIP or AES, direct decryption is possible in theory, but in practice, hacking cases have not been encountered [5,6].

Of course, you can try to pick up the PSK key, or the password to one of the EAP methods. Common attacks on these methods are not known. You can try to apply methods of social engineering, or thermal rectal cryptanalysis.

User can access the network protected by EAP-FAST, EAP-TTLS, PEAP-MSCHAPv2 only by knowing the user's login-password (hacking itself is not possible). Password-type attacks, or those aimed at vulnerabilities in MSCHAP, are also not possible or difficult due to the fact that the EAP channel "client-server" is protected by the encrypted tunnel.

Access to the network that is closed by PEAP-GTC is possible either when the server tokens is compromised or when the token is stolen along with its password [5].

Access to a network that is closed by EAP-TLS is possible when the user certificate is stolen (along with its private key, of course), or when issuing a valid but dummy certificate. This is possible only with the compromise of the certification center, which in normal companies is protected as the most valuable IT resource [6].

Since all the above methods (except for PEAP-GTC) allow the caching of passwords/certificates, when a mobile device is stolen, the attacker gets full access without unnecessary network issues. As a preventive measure, you can use a full hard disk encryption to request a password when the device is turned on.

### IV. TEST FOR PENETRATION IN WI-FI WITH WPA/WPA2-PSK ENCRYPTION

WPA / WPA2 PSK encryptions are vulnerable to dictionary attacks [7]. To implement this attack, you need to obtain a four-way WPA connection between the Wi-Fi client and the access point (AP).

1. Choose the network we need, see what channel it is and fix it in it. After that, we observe what happens in the network.

2. We collect packages Data Packets and Crypt Packets. Their number should increase. If they are equal to 0, then either the network does not have clients, or the network card has not gone into illegible mode.

3. If the packages have gone, then we are waiting for the appearance in the desired network Handshake.

4. Kismet program (forms the Passive WPA Handhsake Collection - Kismet can now collect handshakes from networks and provide them) adds all the useful files to the folder from which it was launched. From Handshake we take the password.

5. Next, create (or find on the Internet) a file with hash passwords for bruteforce and run it. Also, instead of searching through the dictionary, you can run a search of all the passwords in a row.

6. Troubleshoot the hash passwords.

Duration of brute force depends on the hardware and password complexity: it can last from several hours to several days.

## V. Test for penetration in Wi-Fi with WPA2-Enterprise encryption

A hacker attack called "Man in the middle" (or MITM in abbreviation) is the most serious threat to a properly organized WPA2-Enterprise with security certificates [7,8].

To test for penetration in such a network, we can create a fake Wi-Fi-point with a RADIUS server-and get the login, request and response that MS-CHAPv2 uses. This is enough for further password brute force.

We need Kali and a card that supports the work in the mode of Access Point

1. Start Kali Linux.

2. Connect the Wi-Fi-card via USB-OTG-cable. Launch the NetHunter application.

3. Determine the interface of the connected Wi-Fi card.

4. Configure the SSID of the hacked Wi-Fi network.

5. Specify the buffer in which the received logins and hashes will be sent.

6. Write the intercepted data to a file and run Mana.

7. As soon as the Wi-Fi client is close enough to our access point, it will try to authenticate on it.

8. Stop Mana and check what we caught.

9. Crack the received hashes.

The received accounts can be used for further penetration into the corporate network via Wi-Fi or VPN, as well as to gain access to corporate mail.

As it turned out, you can not always intercept user hashes. Desktop OS (Windows, MacOS, Linux), as well as iOS users are protected best. When you first connect, the OS asks if you trust the certificate that is used by the RADIUS server in this Wi-Fi network. When you substitute a legitimate access point, the OS will ask for trust in the new certificate that uses the RADIUS server. This will happen even if you use a certificate issued by a trusted certification authority (i.e.Thawte, Verisign).

## VI. Proposition for increase security

The maximum Wi-Fi network security is provided only by WPA2-Enterprise and digital security certificates in combination with the EAP-TLS or EAP-TTLS protocol. A certificate is a pre-generated file on the RADIUS server and the client device. The client and the authentication server mutually check these files, thereby ensuring protection against unauthorized connections from other devices and false access points. Protocols EAP-TTL / TTLS are included in the 802.1X standard and use the public key infrastructure (PKI) for data exchange between the client and RADIUS. PKI for authentication uses a secret key (the user knows) and the public key (stored in the certificate, potentially known to everyone). The combination of these keys provides reliable authentication.

Digital certificates must be made for each wireless device. This is a laborious process, therefore certificates are usually used only in Wi-Fi networks, which require maximum protection. At the same time, it can be easily revoke the certificate and lock the client.

Today, WPA2-Enterprise in combination with security certificates provides reliable protection for corporate Wi-Fi networks. With proper configuration and use, hacking such protection is almost impossible "from the street", that is, without physical access to authorized client devices. However, network administrators sometimes make mistakes, which leave intruders "loopholes" for penetration into the network. The problem is complicated by the availability of software for hacking and step-by-step instructions, which even amateurs can use.

The administrator must regularly check network traffic for suspicious activity, including delays in the transmission of packets. In areas where there are critical transactions, it is recommended to install Wi-Fi sensors to detect hacking activity in real time.

A special place in the prevention of MITM is the refusal to use ssl-bump filtering. It is often used in offices to prohibit access to certain sites (social networks, entertainment resources, etc.).

## Conclusion

In this paper, a test was conducted for the penetration into a private Wi-Fi network and the organization's network, which are protected by the WPA2-PSK and WPA2-Enterprise protocols, respectively. Both penetration tests were successful. Also the proposals to improve the security of Wi-Fi networks are described in the work.